\documentclass[prd,preprint,nofootinbib]{revtex4}

\usepackage{epsf}
\usepackage{amsmath}
\usepackage{graphics}
\usepackage{amsfonts}
\usepackage{amssymb}
\usepackage{latexsym}
\usepackage{color}
\usepackage[dvips]{graphicx}
\input{colordvi.tex}

\begin{document}

\begin{flushright}
UT-11-21\\
RESCEU-26/11
\end{flushright}

\title{
Curvature perturbation from velocity modulation
}

\author{Kazunori Nakayama$^{a}$ and Teruaki Suyama$^{b}$}

\affiliation{
$^a$Department of Physics, University of Tokyo, Bunkyo-ku, Tokyo 113-0033, Japan\\
$^b$Research Center for the Early Universe, University of Tokyo, Bunkyo-ku, Tokyo 113-0033, Japan\\
}

\vskip 1.0cm

\begin{abstract}
We propose a new variant model of the modulated reheating.
If particles have large scale fluctuations on their velocities, or equivalently their Lorentz factors,
the decay rate also fluctuates and the curvature perturbation is induced via their
decay processes in analogy with modulated reheating.
For example, if they are produced nonthermally by the decay of another field 
with its mass fluctuating on large scales, such a situation is realized.
We explicitly calculate the resulting curvature perturbation and non-linearity parameters
and show that the effect of velocity-modulation is not negligible if the particles are 
semi-relativistic at the decay.
\end{abstract}

\maketitle

%%%%%%%%%%%%%%%%%%%%%%%%%%%%%%%%%%%%
\section{Introduction}
%%%%%%%%%%%%%%%%%%%%%%%%%%%%%%%%%%%%

High accuracy measurements of the cosmic microwave background (CMB) anisotropy 
revealed that the cosmological density perturbations obey the nearly scale-invariant power spectrum~\cite{Komatsu:2010fb}.
The inflaton, which drives the inflationary expansion of the very early Universe,
has quantum fluctuations on large scales and it is a prime candidate for the origin of the
nearly scale invariant curvature perturbation~\cite{Liddle:2000cg}.

However, it was recognized that the curvature perturbation can be generated without invoking the 
quantum fluctuation of the inflaton itself.
There may be many scalar fields in the physics beyond the standard model,
and some of which may be light during inflation and obtain quantum fluctuations.
In the curvaton mechanism~\cite{Mollerach:1989hu,Linde:1996gt,Lyth:2001nq,Moroi:2001ct,Enqvist:2001zp}, 
such a light scalar other than the inflaton, called curvaton, is responsible for the curvature perturbation.
Another possibility is to make the inflaton decay rate fluctuate due to another light scalar
having large scale perturbations~\cite{Kofman:2003nx,Dvali:2003em}.
Then the radiation produced by the inflaton decay has curvature perturbation even if the inflaton itself does not
have enough fluctuations.
This is called the modulated reheating scenario.
Mechanisms for generating the curvature perturbation which shares a similar idea to the modulated reheating
have been proposed~\cite{Dvali:2003ar,Vernizzi:2003vs,Kolb:2004jm,Bauer:2005cd,Lyth:2005qk,Suyama:2006rk,
Matsuda:2007tr,Barnaby:2009mc,Matsuda:2009yt,Kohri:2009ac,Langlois:2009jp}.

In this paper we propose a new mechanism to generate the curvature perturbation,
which is a variant type of the modulated reheating scenario.
The idea is that if a particle velocity is fluctuating on large scales, its decay rate also does
since the lifetime of a particle receives a spatially fluctuating Lorentz boost.
Therefore, the modulated reheating is realized even if an intrinsic decay rate does not depend on 
some fluctuating scalar fields as in the original modulated reheating scenario.

As a concrete example, such a large scale fluctuation in the particle velocity is generated
if it is produced nonthermally by the decay of a heavier particle whose mass depends on another light scalar
with large scale fluctuations.\footnote{
	Here the fluctuation of the particle velocity should not be confused with the ``velocity perturbation'' usually 
	used in the cosmological context.
	The former refers to the velocity of each particle while the latter is the average velocity as the fluid.
	In the present paper we are interested in the former.
}
In this setup, modulated reheating occurs at two stages :
one is at the decay of a heavier particle with a fluctuating mass, the other is at the decay of a daughter particle
whose velocity is fluctuating.
The final curvature perturbation receives both of the two contributions, as well as that generated by the inflaton.
We will systematically calculate the resulting curvature perturbation and estimate the non-Gaussianity,
and find that the effect of modulated velocity is important if the velocity at the decay is not 
much smaller than the speed of light.

%%%%%%%%%%%%%%%%%%%%%%%%%%%%%%%%%%%%
\section{Basic idea}
%%%%%%%%%%%%%%%%%%%%%%%%%%%%%%%%%%%%

Let us suppose that a particle $\sigma$ has a decay width of $\bar\Gamma_\sigma$
at the $\sigma$ rest frame and that its velocity is given by $v$ in the laboratory frame.
Due to the time delay effect, the effective decay rate of $\sigma$ in the laboratory
looks like $\Gamma_\sigma = \bar\Gamma_\sigma /\gamma$ where $\gamma = (1-v^2)^{-1/2}$.

Then let us imagine a situation that the $\sigma$ particles are non-interacting
and have a monochromatic velocity distribution within a small patch of the Universe
but on large scales their velocities fluctuate.
More precisely, we assume that the velocities of $\sigma$-particles have large scale fluctuations 
on the decay hypersurface ($H(\vec x)=\Gamma_\sigma(\vec x)$ in the sudden decay approximation
where $H$ is the Hubble parameter).
On this surface, $\sigma$ has a velocity of
\begin{equation}
	v(\vec x) = \bar v + \delta v(\vec x) .
\end{equation}
Finally the $\sigma$-particles decay into radiation.
Since the decay hypersurface does not coincide with the uniform density hypersurface
due to the fluctuated Lorentz factor, the curvature perturbation is induced at the $\sigma$-decay.
In this setup, the curvature perturbation is estimated as
\begin{equation}
	\zeta \sim \Omega_\sigma\frac{\delta \Gamma_\sigma}{\Gamma_\sigma} = 
	-\Omega_\sigma\frac{\delta\gamma}{\gamma}
	= \Omega_\sigma\gamma^2 \bar v \delta v,
\end{equation}
where $\Omega_\sigma$ is the energy density of the $\sigma$ at the decay relative to the total energy density
and $\delta\Gamma_\sigma(\vec x)\equiv\Gamma_\sigma(\vec x) - \Gamma_\sigma$.
Thus the large scale fluctuation on the velocity can be converted to the curvature perturbation
via the modulated reheating.
Obviously, if the $\sigma$ particles are non-relativistic ($v\to 0$), such an effect is negligible.
Also in the ultra-relativistic limit ($v\to 1$), the curvature perturbation is not generated since
the equation of state of the Universe does not change across the $\sigma$-decay.
Therefore, this effect is expected to be important if the $\sigma$ particles are semi-relativistic 
$v\sim \mathcal O(1)$.
If the $\delta v$ is a Gaussian random variable, then the corresponding curvature perturbation is also Guassian
at the linear order in $\delta v$.

This is a generic idea and may have a potential of broad applications.
In the next section we provide a concrete setup which results in the velocity fluctuation
and induces the curvature perturbation.

%%%%%%%%%%%%%%%%%%%%%%%%%%%%%%%%%%%%
\section{A model of velocity modulation and curvature perturbation}
%%%%%%%%%%%%%%%%%%%%%%%%%%%%%%%%%%%%

%%%%%%%%%%%%%%%%%%%%%%%%%%%%%%%%%%%%%%%%%%
\subsection{Outline}
%%%%%%%%%%%%%%%%%%%%%%%%%%%%%%%%%%%%%%%%%%

We consider a situation that non-relativistic scalar condensate of the $\Sigma$ field decays into
relativistic $\sigma$-particles and subsequently $\sigma$-particles
decay into radiation\footnote{The assumption that $\Sigma$ is a scalar field is not essential to achieve
the velocity modulation. The mechanism can work for the cases where $\Sigma$ is not a scalar field as well.}.
Our key assumption is that the mass of $\Sigma$ spatially fluctuates.
This can be achieved if the mass is dependent on a light scalar field which 
acquired spatial fluctuations during inflation.
Such fluctuations of the mass result in the generation of the primordial curvature 
perturbations through two different processes. 
In the following, let us roughly estimate the resulting curvature perturbation.
More detailed calculations will be given later.

If the mass of $\Sigma$ spatially fluctuates, then generically the decay rate does too.
Let us suppose that $\sigma$ is a fermion.
The $\Sigma$ decays into a $\sigma$ pair through the yukawa interaction, 
\begin{equation}
	\mathcal L = y \Sigma \sigma \bar\sigma,
\end{equation}
where $y$ is the yukawa coupling constant, then the decay rate is given by
\begin{equation}
	\Gamma (\Sigma \to \sigma \bar\sigma) \equiv \Gamma_\Sigma=\frac{y^2}{8\pi} m_\Sigma. \label{yukawa}
\end{equation}
If $\sigma$ is a scalar, $\Sigma$ can decay into them through the three point interaction,
\begin{equation}
	\mathcal L = \mu \Sigma \sigma \sigma,
\end{equation}
where $\mu$ is a coupling constant having mass dimension one. 
Then the decay rate is given by
\begin{equation}
	\Gamma (\Sigma \to \sigma \sigma) \equiv \Gamma_\Sigma=\frac{\mu^2}{8\pi m_\Sigma} \label{three}.
\end{equation}
It is clear from these equations that fluctuation of $m_\Sigma$ is taken over by $\Gamma_\Sigma$.
%Following arguments do not depend on whether $\sigma$ is a scalar or fermion.
For concreteness we assume $\sigma$ is a fermion in the following, 
although qualitative arguments do not depend on whether it is a fermion or scalar.
The density fluctuations of $\sigma$ are produced when $\Sigma$ decays;
\begin{equation}
	\frac{\delta \rho_\sigma}{\rho_\sigma} \simeq \frac{\delta m_\Sigma}{m_\Sigma}.
\end{equation}
This is exactly the same mechanism of generating the curvature perturbation
in the modulated reheating scenario.

Since each $\sigma$-particle has an energy $m_\Sigma/2$ when it is produced,
velocity of the produced $\sigma$-particle is given by
\begin{equation}
v_0=\frac{\sqrt{m_\Sigma^2-4m_\sigma^2}}{m_\Sigma}.
\end{equation}
It is clear that the velocity $v$ also fluctuates if $m_\Sigma$ fluctuates.
We assume that the interaction of $\sigma$-particle is sufficiently weak so that it
does not take part in thermal bath and the velocity remains constant 
except for the redshift by the cosmic expansion.\footnote{
	The $\sigma$-particles can scatter off themselves by the exchange of $\Sigma$.
	This process is suppressed if the coupling constant between $\Sigma$ and $\sigma$
	is small and/or the $\Sigma$ is heavy enough.
}
Otherwise, the $\sigma$-particles obey thermal distribution with a background temperature 
of the radiation and the velocity fluctuation will be smoothed out as long as $\sigma$ is subdominant.
Now, let us suppose that the velocity of $\sigma$-particle has dropped to $v_1$
due to the cosmological redshift when $\sigma$-particles decay.
Taking into account the effect of time delation, the decay rate of $\sigma$ is given by
\begin{equation}
\Gamma (\sigma \to {\rm radiation}) \equiv \Gamma_\sigma = {\bar \Gamma_\sigma} \sqrt{1-v_1^2},
\end{equation}
where ${\bar \Gamma_\sigma}$ is the decay rate measured at the rest frame.
Since $v_1$ is position dependent, this equation manifests that the decay rate of 
$\sigma$ spatially fluctuates too;
\begin{equation}
\frac{\delta \Gamma_\sigma}{\Gamma_\sigma} \simeq v_1^2 \frac{\delta m_\Sigma}{m_\Sigma}.
\end{equation}
Again, the modulation of $\Gamma_\sigma$ gives additional contribution
to the curvature perturbation.
Denoting by $\Omega_\sigma$ a fraction of $\sigma$-particles when they decay,
we expect that the final curvature perturbation $\zeta$ is given by
\begin{equation}
\zeta \simeq \Omega_\sigma \frac{\delta \rho_\sigma}{\rho_\sigma}+\Omega_\sigma \frac{\delta \Gamma_\sigma}{\Gamma_\sigma} \simeq \Omega_\sigma (1+v_1^2) \frac{\delta m_\Sigma}{m_\Sigma}. \label{naivezeta}
\end{equation}
Therefore, if $\sigma$-particles are still relativistic when they decay,
i.e. $v_1={\cal O}(1)$, then the effect of the velocity modulation on the final
curvature perturbation cannot be neglected.

%%%%%%%%%%%%%%%%%%%%%%%%%%%%%%%%%%%%
\subsection{Calculation of the curvature perturbation}
%%%%%%%%%%%%%%%%%%%%%%%%%%%%%%%%%%%%

Now let us evaluate the curvature perturbation based on the $\delta N$-formalism
\cite{Sasaki:1995aw,Lyth:2004gb}.
We adopt the so-called sudden decay approximation in which the decay 
of $\Sigma$ or $\sigma$-particles is assumed to occur instantaneously when the decay
rate becomes equal to the Hubble expansion rate~\cite{Sasaki:2006kq}.

Let us take a hypersurface just before the $\Sigma$ field decays on 
which the total energy density is spatially uniform (uniform density hypersurface).
We here assume that $\Sigma$ begins to oscillate in a period between the end of inflation and
the completion of reheating where the universe expands like a matter-dominated universe
and that the initial amplitude is constant.
In this case the energy density of the $\Sigma$ condensation does not fluctuate on the uniform density slice
even if its mass fluctuates.\footnote{
	If, otherwise, the $\Sigma$ begins to oscillate during the radiation-dominated era after the reheating,
	the energy density of $\Sigma$ itself also has fluctuations.
	We do not consider such a case since it only makes the following analyses more complicated.
}
Since no fluctuations exist before $\Sigma$ decays, this hypersurface
coincides with the spatially flat hypersurface.
Let us denote by $\delta N_1$ a required e-folding number from the flat hypersurface
to the decay hypersurface on which $\Sigma$ decays into $\sigma$-particles.
Since $\Gamma_\Sigma$ spatially fluctuates, $\delta N_1$ does too.
Thus we have
\begin{equation}
	\bar \rho_r e^{-4\delta N_1} + \bar \rho_\Sigma e^{-3\delta N_1} = \bar \rho_{\rm tot}
	\left( 1+ \frac{\delta \Gamma_\Sigma}{\Gamma_\Sigma} \right)^2 ,
\end{equation}
where $\rho_r$ and $\rho_\Sigma$ denote the energy densities of the radiation and $\Sigma$ at the $\Sigma$-decay.
The total energy density is given by $\rho_{\rm tot} = \rho_r+\rho_\Sigma$.
Quantities with bars represent the background values.
This is rewritten in the form as
\begin{equation}
(1-\Omega_\Sigma) e^{-4 \delta N_1}+\Omega_\Sigma e^{-3 \delta N_1}={\left( 1+\frac{\delta \Gamma_\Sigma}{\Gamma_\Sigma} \right)}^2. \label{deltaN1}
\end{equation}
where $\Omega_\Sigma \equiv \bar\rho_\Sigma/\bar\rho_{\rm tot}$ at the $\Sigma$-decay.
Let us next consider a hypersurface just after the $\Sigma$ decays
on which $\rho_\sigma$ is spatially uniform and denote by $\delta N_2$
a required e-folding number from the decay hypersurface to uniform $\rho_\sigma$ hypersurface. 
Then equation for $\delta N_2$ is given by
\begin{equation}
e^{-3 \delta N_1} e^{-3 (1+w_0) \delta N_2}=1, \label{deltaN2}
\end{equation}
where $w_0$ is the effective equation of state parameter $w_0 = P_\sigma/\rho_\sigma$
of $\sigma$ evaluated on the decay surface.
According to the $\delta N$ formalism, the sum of $\delta N_1$ and $\delta N_2$
gives the curvature perturbation of the uniform $\rho_\sigma$ hypersurface;
\begin{equation}
\zeta_\sigma = \delta N_1+\delta N_2. \label{zetasigma}
\end{equation}
Eqs.~(\ref{deltaN1}), (\ref{deltaN2}) and (\ref{zetasigma}) allow us
to express $\zeta_\sigma$ in terms of $\delta \Gamma_\Sigma$, which we defer.
Unless $w_0$ is zero, $\zeta_\sigma$ is non-vanishing, as it should be.
Since $\sigma$-particles do not interact with the other particles,
$\zeta_\sigma$ is conserved until $\sigma$-particles decay into radiation.
On the other hand, $\zeta_r$, the curvature perturbation on the uniform $\rho_r$ hypersurface, is zero since we assume that $\Sigma$
decays only into $\sigma$-particles.

Now we consider a difference between a hypersurface on which $\sigma$-particles 
decay into radiation and a one on which $\rho_\sigma$ is uniform.
Let us denote by $\delta N_3$ a required e-folding number from uniform 
$\rho_\sigma$ hypersurface to $\sigma$ decay hypersurface.
Then we obtain
\begin{equation}
	\bar\rho_\sigma e^{-3(1+w_1)\delta N_3} + \bar\rho_r e^{-4(\zeta_\sigma + \delta N_3)} = 
	\bar \rho_{\rm tot} \left( 1+ \frac{\delta \Gamma_\sigma}{\Gamma_\sigma} \right)^2.
\end{equation}
This is rewritten as
\begin{equation}
\Omega_\sigma e^{-3 (1+w_1) \delta N_3}+(1-\Omega_\sigma) e^{-4(\zeta_\sigma+\delta N_3)}={\left( 1+\frac{\delta \Gamma_\sigma}{\Gamma_\sigma} \right)}^2, \label{deltaN3}
\end{equation}
where $\Omega_\sigma \equiv \bar\rho_\sigma / \bar \rho_{\rm tot}$ on the $\sigma$-decay surface and
$w_1$ is the effective equation of state parameter of $\sigma$ on this hypersurface.
Let us next consider a hypersurface just after the $\sigma$-particles decay
on which radiation energy density is spatially uniform and denote by $\delta N_4$
a required e-folding number from the $\sigma$ decay hypersurface to 
the uniform density hypersurface. 
Then we have the following relation
\begin{equation}
e^{4 \delta N_4}={\left( 1+\frac{\delta \Gamma_\sigma}{\Gamma_\sigma} \right)}^2. \label{deltaN4}
\end{equation}
The final curvature perturbation is given by
\begin{equation}
\zeta=\zeta_\sigma+\delta N_3+\delta N_4.
\end{equation}
From Eqs.~(\ref{deltaN3}) and (\ref{deltaN4}),
we can express both $\delta N_3$ and $\delta N_4$ as a function
of $\delta \Gamma_\sigma$.
Therefore, $\zeta$ can be written in terms of $\delta \Gamma_\Sigma$
and $\delta \Gamma_\sigma$.
Note that $\Omega_\Sigma$ and $\Omega_\sigma$ are related through
\begin{equation}
	\Omega_\sigma = \frac{\Omega_\Sigma}{\Omega_\Sigma+(1-\Omega_\Sigma)
	\exp \left[ \int 3H(1+w(t))dt - 4N \right]},    \label{Os-OS}
\end{equation}
where the e-folding number $N$ measures the duration between the
$\Sigma$ and $\sigma$ decay hypersurfaces
and $w(t)$ denotes the time-dependent equation of state of the $\sigma$ particle.
The integral in the exponent starts from the time of $\Sigma$-decay and ends at the $\sigma$-decay.
It is soon seen that in the relativistic limit $w=1/3$, $\Omega_\sigma$ remains constant 
as is expected.
The equation of state of $\sigma$ at its production $(w_0)$ and decay $(w_1)$ are also related.
First note that the equation of state of the $\sigma$ is given by
\begin{equation}
	w(t) = \frac{P_\sigma}{\rho_\sigma} = \frac{p^2(t)}{3(p^2(t)+m_\sigma^2)},
\end{equation}
because it has a monochromatic momentum distribution in the sudden decay approximation,
where $p(t)$ is the momentum of the $\sigma$ particle.
Then we obtain
\begin{equation}
	w_1=\frac{w_0}{(1-3w_0)e^{2N}+3w_0}. \label{w1}
\end{equation}
Since each $\sigma$-particle has an energy of $m_\Sigma/2$ at the time
of their creation, $w_0$ can be written as
\begin{equation}
	w_0=\frac{m_\Sigma^2-4m_\sigma^2}{3m_\Sigma^2}. \label{w0}
\end{equation}

To get the final expression for the curvature perturbation, we need to express $\delta \Gamma_\Sigma$
and $\delta \Gamma_\sigma$ as the functions of $\delta m_\Sigma$.
As for the former, it is already given by Eq.~(\ref{yukawa}).
%As for the latter, we start from noting that $\Gamma_\sigma$ in terms of
%$w = P_\sigma/ \rho_\sigma$ is written as
%\begin{equation}
%\Gamma_\sigma={\bar \Gamma_\sigma} \sqrt{1-3w_1}. \label{Gammas}
%\end{equation}
%The evolution equation for $w$ is given by
%\begin{equation}
%\frac{dw}{dN}=-2w (1-3w).
%\end{equation}
%By solving this equation, we find that $w_0$ and $w_1$ are related by
As for the latter, the decay rate of $\sigma$ is calculated from
\begin{equation}
	\Gamma_\sigma={\bar \Gamma_\sigma} \sqrt{1-3w_1}. \label{Gammas}
\end{equation}
Through Eqs.~(\ref{Gammas}), (\ref{w1}) and (\ref{w0}), we see that
$\Gamma_\sigma$ is related to $m_\Sigma$.
To third order in $\delta m_\Sigma$, we find
\begin{equation}
\frac{\delta \Gamma_\sigma}{\Gamma_\sigma}=-\frac{w_1}{w_0} \frac{\delta m_\Sigma}{m_\Sigma}-\frac{(w_0-3 w_1)w_1}{2w_0^2}  {\left( \frac{\delta m_\Sigma}{m_\Sigma} \right)}^2+\frac{(3w_0-5w_1)w_1^2}{2w_0^3} {\left( \frac{\delta m_\Sigma}{m_\Sigma} \right)}^3.
\end{equation}

Using these results obtained above, we can Taylor-expand $\zeta$ in terms of $\delta m_\Sigma$ to any order.
To third order, it is given by
\begin{equation}
	\zeta=A_1 \frac{\delta m_\Sigma}{m_\Sigma}
	+\frac{1}{2} A_2 {\left( \frac{\delta m_\Sigma}{m_\Sigma} \right)}^2+\frac{1}{6}A_3  {\left( \frac{\delta m_\Sigma}{m_\Sigma} \right)}^3, \label{finalzeta}
\end{equation}
where $A_1$ is given by
\begin{equation}
	A_1=\frac{\Omega_\sigma  \left[12 w_0^2 (w_1+1)-w_1 (w_0+1) (3 w_1-1) (\Omega_\Sigma -4)\right]}
	{2 w_0 (w_0+1) (\Omega_\Sigma -4) ((3 w_1-1) \Omega_\sigma +4)}. \label{coeA1}
\end{equation}
The other expansion coefficients $A_2$ and $A_3$ are given in the appendix.
Let us confirm that apart from the ${\cal O}(1)$ numerical factors, 
Eq.~(\ref{coeA1}) reproduces the naive expectation Eq.~(\ref{naivezeta}).
First, we see that $A_1$ is proportional to $\Omega_\sigma$ (It can be confirmed that
$A_2$ and $A_3$ are also proportional to $\Omega_\sigma$.).
Therefore, $\zeta$ of Eq.~(\ref{finalzeta}) is proportional to $\Omega_\sigma$ and vanishes
in the limit $\Omega_\sigma \to 0$, which is also true for the naive expectation.
Secondly, let us expand $A_1$ in terms of $w_1$;
\begin{equation}
A_1=\frac{3 \Omega_\sigma }{8 (\Omega_\sigma -4)}+\frac{9 \Omega_\sigma }{2 (\Omega_\sigma -4)^2} w_1
+{\cal O}(w_1^2), \label{expandA1}
\end{equation}
where we have set $\Omega_\Sigma=0,~w_0=1/3$ for simplicity.\footnote{
	Actually $\Omega_\sigma$ and $\Omega_\Sigma$ are related through Eq.~(\ref{Os-OS}).
	We can always take $\Omega_\Sigma \ll \Omega_\sigma$ by choosing $N$ and $w_0$.
}
We see that the first term remains even if we set $w_1=0$.
Therefore, this term represents the contribution of $\delta \rho_\sigma$ in Eq.~(\ref{naivezeta}), 
which originates from the modulation of $\Gamma_\Sigma$.
On the other hand, the second term in Eq.~(\ref{expandA1}) is proportional
to $w_1$ and hence proportional to $v_1^2$.
This corresponds to the second term in Eq.~(\ref{naivezeta}) and is due to
the modulation of $\Gamma_\sigma$.

From Eq.~(\ref{finalzeta}), the scalar spectral index, $n_s$, is calculated as~\cite{Ichikawa:2008ne}
\begin{equation}
	n_s = 1-2\epsilon - \frac{4\epsilon - 2\eta}{1+ 2 M_P^2 \epsilon N_\chi^2 },
\end{equation}
where we have defined $N_\chi$ through $A_1 \delta m_\Sigma / m_\Sigma \equiv N_\chi \delta \chi$
with $\chi$ being the light field giving spatial modulation to the $\Sigma$ mass,
$M_P$ denotes the reduced Planck scale,
and $\epsilon$ and $\eta$ are inflationary slow-roll parameters~\cite{Liddle:2000cg}.
If the inflaton dominantly contributes to the curvature perturbation, we recover the standard formula, $n_s = 1-6\epsilon + 2\eta$.
Otherwise, it approaches to $n_s = 1-2\epsilon$, similarly to the curvaton case.

%%%%%%%%%%%%%%%%%%%%%%%%%%%%%%%%%%%%
\subsection{Non-linearity parameters}
%%%%%%%%%%%%%%%%%%%%%%%%%%%%%%%%%%%%

\begin{figure}[t]
  \begin{center}{
    \includegraphics[scale=0.7]{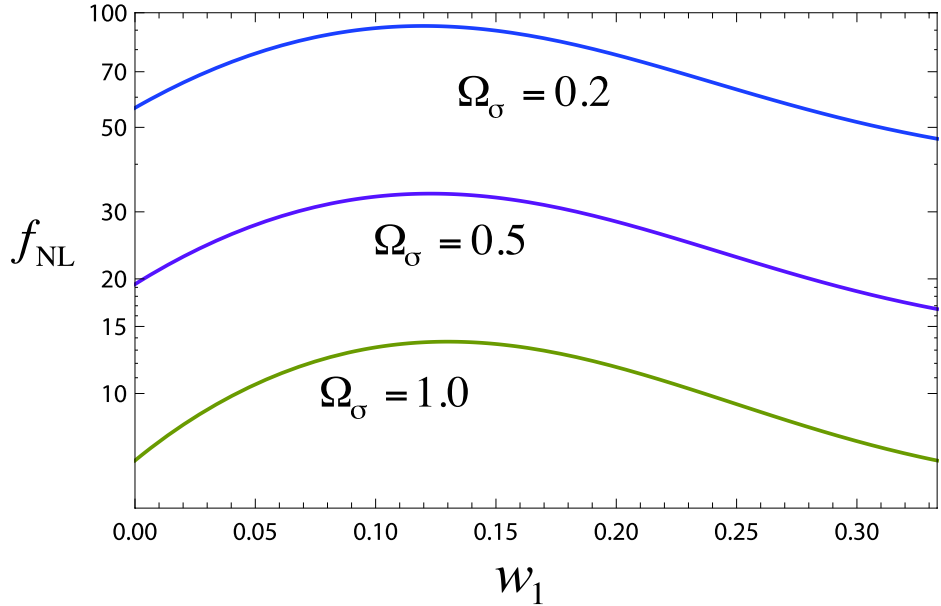}
    }
  \end{center}
  \caption{Plots of $f_{\rm NL}$ as a function of $w_1$ for three cases $\Omega_\sigma=(0.2,~0.5,~1.0)$.}
 \label{non-linear}
\end{figure}

\begin{figure}[t]
  \begin{center}{
    \includegraphics[scale=0.7]{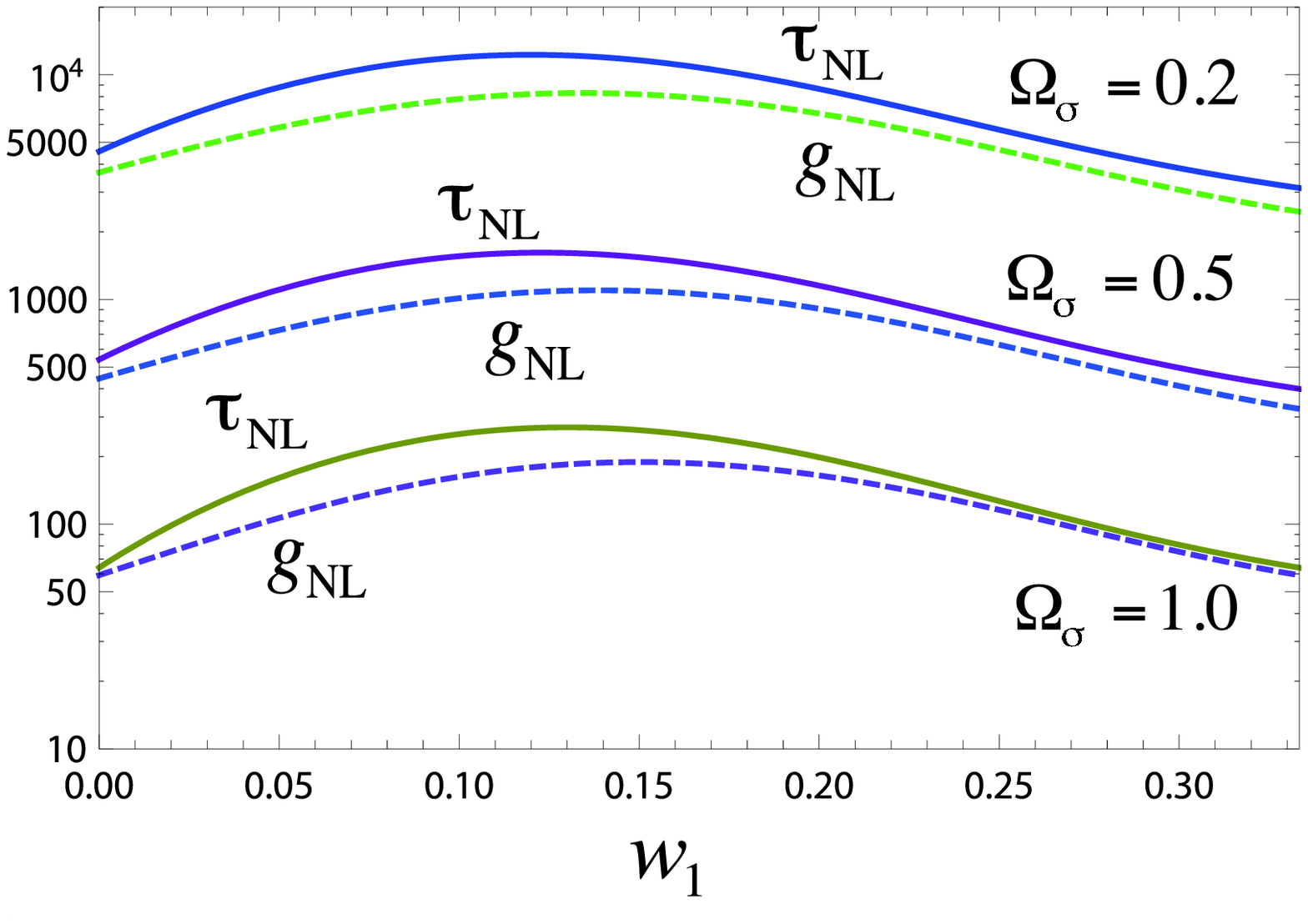}
    }
  \end{center}
  \caption{Plots of $\tau_{\rm NL}$ and $g_{\rm NL}$ as a function of $w_1$ for three cases $\Omega_\sigma=(0.2,~0.5,~1.0)$.}
 \label{non-linear-gnl}
\end{figure}

In this paper, we consider the simplest case where $\delta m_\Sigma$ is a gaussian variable.
For simplicity, let us assume that $\Sigma$ is subdominant when they decay,
i.e., $\Omega_\Sigma \ll 1$ and $\sigma$-particles are relativistic when they are created, i.e., $w_0 = 1/3$.
Hereafter we neglect the contribution of the inflaton fluctuation to the curvature perturbation
and the dominant contribution to the curvature perturbation comes from the fluctuation of $m_\Sigma$.
With these assumptions, the non-linearity parameters are given by\footnote{
	The non-linearity parameters for the standard modulated reheating case
	are given in Ref.~\cite{Suyama:2007bg}.
}
\begin{eqnarray}
f_{\rm NL}=&&\frac{20}{9 (3 w_1 (4 w_1-1)+1)^2 \Omega_\sigma  ((3
   w_1-1) \Omega_\sigma +4)}  \nonumber \\
   && \times \bigg[ w_1 (36 w_1 (2 w_1-1)+5)+1) (\Omega_\sigma -3 w_1 \Omega_\sigma )^2 \nonumber \\
   &&+2 (w_1 (3 w_1 (24 w_1 (6 w_1-7)+35)+10)-7) \Omega_\sigma \nonumber \\
   &&+2 w_1 (9 w_1 (8 w_1 (6 w_1+5)-21)+22)+22 \bigg], \label{fnl} 
\end{eqnarray}
\begin{eqnarray}
\tau_{\rm NL}=&&\frac{36}{25} f_{\rm NL}^2, \label{taunl} 
\end{eqnarray}
\begin{eqnarray}
g_{\rm NL}=&&\frac{800}{243 (3 w_1 (4 w_1-1)+1)^3 \Omega_\sigma ^2 (-3
   w_1 \Omega_\sigma +\Omega_\sigma -4)^2}   \nonumber \\
 \times &&\bigg[ (3 w_1-1) (w_1 (3 w_1 (27 w_1 (4 w_1 (96 w_1 (2
   w_1-3)+127)-75)-43)-97)+41) \Omega_\sigma ^3  \nonumber \\ 
   &&+3 (w_1 (3 w_1 (9 w_1 (w_1 (12
   w_1 (8 w_1 (36 w_1+19)-413)+2579)-416)+262)-500)+133) \Omega_\sigma ^2 \nonumber \\
   &&+2 (w_1 (36
   w_1 (3 w_1-1) (4 w_1-1)+1)+1) (\Omega_\sigma -3 w_1 \Omega_\sigma )^4 \nonumber \\
   &&+2 (w_1 (3
   w_1 (3 w_1 (3 w_1 (12 w_1 (12 (37-12 w_1) w_1+133)-5177)+4570)-728)+674)-517) \Omega_\sigma \nonumber \\
   && +4 (w_1+1) (9 w_1 (w_1 (12 w_1 (36
   w_1 (4 w_1+9)-181)+265)-2)+209)\bigg]. \label{gnl}
\end{eqnarray}
In Fig.~\ref{non-linear} and \ref{non-linear-gnl}, we show plots of $f_{\rm NL},~\tau_{\rm NL}$ and $g_{\rm NL}$
as a function of $w_1$ for three cases $\Omega_\sigma=(0.2,~0.5,~1.0)$.
From this, we see that $f_{\rm NL}$ is always positive.
For fixed $w_1$, the smaller $\Omega_\sigma$ is, the larger $f_{\rm NL}$ is.
Actually, we can easily show from Eq.~(\ref{fnl}) that $f_{\rm NL} \simeq \Omega_\sigma^{-1}$ 
when $\Omega_\sigma$ is small.
The mechanism of this boost of $f_{\rm NL}$ is exactly the same as that in the curvaton model
in which $f_{\rm NL}$ is inversely proportional to curvaton fraction evaluated
at the time when the curvaton decays. 
For fixed $\Omega_\sigma$, $f_{\rm NL}$ has a maximum at $w_1 \simeq 0.1$
and that the maximum value is roughly two times larger than $f_{\rm NL}$ at $w_1=0$.
Therefore, if the motion of $\sigma$-particles are mildly relativistic,
then the effect of the velocity modulation can double $f_{\rm NL}$.

We also find that $g_{\rm NL}$ is always positive and
$\tau_{\rm NL}$ is always larger than $g_{\rm NL}$.
Noticing that their difference is less than a factor 2 and using Eq.~(\ref{taunl}),
we have $g_{\rm NL} = {\cal O}(f_{\rm NL}^2)$.
Therefore, in this model,
both $\tau_{\rm NL}$ and $g_{\rm NL}$ become large when $f_{\rm NL}$ is large.

%%%%%%%%%%%%%%%%%%%%%%%%%%%%%%%%%%%%
\section{Conclusions and discussion}
%%%%%%%%%%%%%%%%%%%%%%%%%%%%%%%%%%%%

We have shown that the curvature perturbation can be sourced by the decay of particles with
velocity modulation.
We have given a concrete setup for realizing such a scenario.
A class of modulated reheating scenario in which the mass of decaying particle fluctuates
generates a velocity modulation of daughter particles, 
which may result in the additional source of curvature perturbation.
The non-Gaussian signatures may also be enhanced due to the velocity-modulation effects.

One can easily construct a variant model.
For example, let us suppose that the mass of $\Sigma$ does not fluctuate but the $\sigma$-mass does.
The velocity of $\sigma$, produced by the $\Sigma$-decay,
also fluctuates in this case and the velocity modulation can generate
the curvature perturbation at the $\sigma$-decay in a same manner.

The same mechanism may generate the CDM/baryon isocurvature perturbation
if the $\sigma$-decay produces CDM/baryon.
Actually $\sigma$ can be identified with the right-handed neutrino $N_R$
while $\Sigma$ with a scalar field giving the mass for $N_R$.
The relevant interaction Lagrangian is given by
\begin{equation}
	\mathcal L = \left(y_i \Sigma N_{Ri}\bar N_{Ri}^c + y_{ij}^{(\nu)} H L_i \bar N_{Rj} + {\rm h.c.}\right) - V(\Sigma),
\end{equation}
where $i,j$ are generation indices and the scalar potential of $\Sigma$ is taken to be
\begin{equation}
	V(\Sigma) =  -\mu_\Sigma^2 |\Sigma|^2 + \lambda |\Sigma|^4.
\end{equation}
Here $H$ is the standard model Higgs boson and $L_i$ is the lepton doublet.
The $\Sigma$ mass around the vacuum,
$m_{\Sigma}=2\mu_\Sigma$, is assumed to depend on another light scalar having large scale fluctuations.
This model possesses a global U(1)$_L$ or gauged U(1)$_{B-L}$ symmetry 
which is spontaneously broken by the VEV of $\Sigma$.\footnote{
	In the case of global symmetry, there exists a Goldstone mode, called the majoron~\cite{Chikashige:1980ui}.
	It does not have significant cosmological and astrophysical effects if the breaking scale is
	sufficiently large.
}
In this model $\Sigma$ decays into the $N_R$-pair, and subsequently $N_R$ decays into the
Higgs boson and lepton, generating the lepton asymmetry (which is converted to the baryon asymmetry
through the sphaleron process) if the CP angle is nonzero~\cite{Fukugita:1986hr}.
Since $N_R$ has a large scale velocity modulation, the baryon number created by its decay
also has fluctuations.
If the $\Sigma$ or $N_R$ decay gives dominant curvature perturbation, there is no baryonic isocurvature perturbation.
On the other hand, if the curvature perturbation is dominantly sourced by the inflaton fluctuation,
this results in the baryonic isocurvature perturbation.

%%%%%%%%%%%%%%%%%%%%%%%%%%%%%%%%%%%%
\section*{Acknowledgment}
%%%%%%%%%%%%%%%%%%%%%%%%%%%%%%%%%%%%

This work is supported by Grant-in-Aid for Scientific research from
the Ministry of Education, Science, Sports, and Culture (MEXT), Japan,
No.\ 21111006 (K.N.) and No.\ 22244030 (K.N.).
and by Grant-in-Aid for JSPS Fellows No.~1008477 (T.S.).
TS thanks the Centre for Cosmology, Particle Physics and Phenomenology
at Universit\'{e} catholique de Louvain for its hospitality during the completion of this work.

%%%%%%%%%%%%%%%%%%%%%%%%%%%%%%%%%%%%%%%%%%%%
\appendix
%%%%%%%%%%%%%%%%%%%%%%%%%%%%%%%%%%%%%%%%%%%%

\section{Expressions for $A_2$ and $A_3$}

Here we give expressions for $A_2$ and $A_3$ in Eq.~(\ref{finalzeta}).

\begin{eqnarray}
A_2=&&\frac{1}{12 w_0^5 (w_1+1)^2 \Omega_\sigma ^2 ((3 w_1-1) \Omega_\sigma +4)} \bigg[ 8 A_1^3 w_0^3 (w_0+1)^2 ((3 w_1-1) \Omega_\sigma +4)^3 \nonumber \\
&& +4 A_1^2 w_0^2 \Omega_\sigma  \left(w_0^3 (w_1+1) \left(63 w_1^2 \Omega_\sigma ^2-6 w_1 \left(7 \Omega
   \sigma ^2-20 \Omega_\sigma -8\right)+7 \Omega_\sigma ^2-104 \Omega_\sigma +160\right) \right. \nonumber \\
   && \left. +w_0^2 \left(9 w_1^2+4 w_1+7\right) (-3 w_1 \Omega_\sigma +\Omega_\sigma -4)^2+6 w_0 w_1 (3
   w_1-1) (-3 w_1 \Omega_\sigma +\Omega_\sigma -4)^2 \right. \nonumber \\
   && \left. +3 w_1 (3 w_1-1) (-3 w_1 \Omega_\sigma +\Omega_\sigma -4)^2\right)+2 A_1 w_0 \Omega_\sigma ^2 \left(6 w_0^4
   (w_1+1)^2+14 w_0^3 w_1 \left(3 w_1^2+2 w_1-1\right) \right. \nonumber \\
   && \left. +w_0^2 w_1 \left(27 w_1^3+24 w_1^2+31 w_1-14\right)+6 w_0 (1-3
   w_1)^2 w_1^2+3 (1-3 w_1)^2 w_1^2\right) ((3 w_1-1) \Omega_\sigma +4) \nonumber \\
   &&+w_1^2 (3 w_1-1) \Omega_\sigma ^3 \left(w_0^3 \left(33 w_1^2+38
   w_1+5\right)+w_0^2 \left(9 w_1^3+15 w_1^2+15 w_1-7\right)+2 w_0 (1-3 w_1)^2 w_1 \right. \nonumber \\
   && \left. +(1-3 w_1)^2 w_1\right) \bigg], \\
A_3=&&A_{3,1}+A_{3,2}+A_{3,3},
\end{eqnarray}
where 
\begin{eqnarray}
A_{3,1}=&&\frac{3 (3 w_1-1) w_1 (w_0-3 w_1) (16 A_1 w_0 (\Omega_\sigma -1)+w_1 \Omega_\sigma  (-3 w_1 \Omega_\sigma +\Omega_\sigma -4))}{2 w_0^3 ((3 w_1-1)
   \Omega_\sigma +4)^2} \nonumber \\
   &&+\frac{w_0 f_1}{w_0+1}-\frac{w_1^3}{w_0^3}+\frac{3 (3 w_1-1) w_1^2 \Omega_\sigma  (3 w_0-5 w_1)}{2 w_0^3 ((3 w_1-1) \Omega_\sigma +4)}, \\
A_{3,2}=&&\frac{2}{3 w_0^6 (w_0+1) (w_1+1)^4 \Omega_\sigma ^4 ((3 w_1-1) \Omega_\sigma
   +4)} \bigg[ 12 (w_0+1) w_0^3 (w_1+1)^4 (\Omega_\sigma -1) \Omega_\sigma ^4 (2 A_1 w_0+w_1)^3 \nonumber \\
   &&-12 (w_0+1) w_0^3 (w_1+1)^4 \Omega_\sigma ^2 (2 A_1 w_0 (\Omega_\sigma
   -1)+w_1 \Omega_\sigma )^3+6 f_1 w_0^7 (\Omega_\sigma -1) (w_1 \Omega_\sigma +\Omega_\sigma )^4\bigg], \\
A_{3,3}=&&\frac{2}{3 w_0^6 (w_0+1) (w_1+1)^4 \Omega_\sigma ^4 ((3 w_1-1) \Omega_\sigma +4)}
 \bigg[ (w_0+1) (w_1+1) (1-\Omega_\sigma ) \Omega_\sigma  (2 A_1 w_0+w_1)  \nonumber \\
 &&  \bigg \{ (2 A_1 w_0 ((3 w_1-1) \Omega_\sigma +4)+w_1 (3 w_1-1) \Omega_\sigma )
   \left(2 A_1 w_0 (w_0+1) ((3 w_1-1) \Omega_\sigma +4)  \right. \nonumber \\
   && \left. +\Omega_\sigma  \left(w_0^2 (w_1+1)+w_0 w_1 (3 w_1-1)+w_1 (3 w_1-1)\right)\right)
   \left(2 A_1 w_0 (w_0+1) ((3 w_1-1) \Omega_\sigma +4) \right. \nonumber \\
   && \left.+\Omega_\sigma  \left(6 w_0^2 (w_1+1)+w_0 w_1 (3 w_1-1)+w_1 (3 w_1  -1)\right)\right)-\frac{72 f_2 w_0^5 (w_1+1)^3 \Omega_\sigma ^3}{(3 w_1-1) \Omega_\sigma +4} \bigg \} \nonumber \\
   &&-\frac{54 f_2 w_0^5 (w_0+1) (w_1+1)^5 \Omega_\sigma ^4 (2 A_1 w_0
   (\Omega_\sigma -1)+w_1 \Omega_\sigma )}{(3 w_1-1) \Omega_\sigma +4} \bigg],
\end{eqnarray}
and $f_1$ and $f_2$ are defined by
\begin{eqnarray}
f_1=&&\frac{3 w_1^2 (3 w_1-1) \Omega_\sigma  (3 w_0-5 w_1)}{2 w_0^3 ((3
   w_1-1) \Omega_\sigma +4)}, \\
f_2=&&\frac{3 w_1 (3 w_1-1) (w_0-3 w_1) (16 A_1 w_0 (\Omega_\sigma
   -1)+w_1 \Omega_\sigma  (-3 w_1 \Omega_\sigma +\Omega_\sigma -4))}{2 w_0^3 (-3 w_1
   \Omega_\sigma +\Omega_\sigma -4)^2}.
\end{eqnarray}

%%%%%%%%%%%%%%%%%%%%%%%%%%%%%%%%%%%%%%%%%%%%
  
%%%%%%%%%%%%%%%%%%%%%%%%%%%%%%%%%%%%%%%%%%%%%

\end{document}